\documentclass[10pt, conference, compsocconf]{IEEEtran}
\usepackage[utf8]{inputenc}
\usepackage{balance}
\usepackage{graphicx}
\usepackage{todonotes}
\usepackage{acro}
\usepackage{url}
\usepackage{textcomp}

\DeclareAcronym{cps}{
  short = CPS,
  long  = Cyber Physical System,
  sort  = C,
}
\DeclareAcronym{lec}{
  short = LEC,
  long  = Learning Enabled Component,
  sort  = L,
}
\DeclareAcronym{uuv}{
  short = UUV,
  long  = Unmanned Underwater Vehicle,
  sort  = U,
}
\DeclareAcronym{cnn}{
  short = CNN,
  long  = Convolutional Neural Network,
  sort  = C,
}
\DeclareAcronym{dsml}{
  short = DSML,
  long  = Domain Specific Modeling Language,
  sort  = D,
}
\DeclareAcronym{gsn}{
  short = GSN,
  long  = Goal Structuring Notation,
  sort  = G,
}
\DeclareAcronym{ros}{
  short = ROS,
  long  = Robot Operating System,
  sort  = R,
}
\DeclareAcronym{alc}{
  short = ALC,
  long  = Assurance-based Learning-enabled CPS,
  sort  = A,
}
\DeclareAcronym{ann}{
  short = ANN,
  long  = Artificial Neural Network,
  sort  = A,
}
\DeclareAcronym{dag}{
  short = DAG,
  long  = Directed Acyclic Graph,
  sort  = D,
}
\DeclareAcronym{ml}{
  short = ML,
  long  = Machine Learning,
  sort  = M,
}

\begin{document}

\title{Workflow Automation for Cyber Physical System Development  Processes}

\author{
    \IEEEauthorblockN{Charles Hartsell\\
    Nagabhushan Mahadevan\\
    Harmon Nine}
    \IEEEauthorblockA{Institute for Software Integrated Sys.\\
    Vanderbilt University\\
    charles.a.hartsell@vanderbilt.edu}
    \and
    \IEEEauthorblockN{Ted Bapty\\
    Abhishek Dubey\\
    Gabor Karsai}
    \IEEEauthorblockA{Institute for Software Integrated Sys.\\
    Vanderbilt University\\
    gabor.karsai@vanderbilt.edu}
}

\date{January 2019}

\maketitle

\begin{abstract}
Development of Cyber Physical Systems (CPSs) requires close interaction between developers with expertise in many domains to achieve ever-increasing demands for improved performance, reduced cost, and more system autonomy. Each engineering discipline commonly relies on domain-specific modeling languages, and analysis and execution of these models is often automated with appropriate tooling. However, integration between these heterogeneous models and tools is often lacking, and most of the burden for inter-operation of these tools is placed on system developers. To address this problem, we introduce a workflow modeling language for the automation of complex CPS development processes and implement a platform for execution of these models in the Assurance-based Learning-enabled CPS (ALC) Toolchain. Several illustrative examples are provided which show how these workflow models are able to automate many time-consuming integration tasks previously performed manually by system developers.
\end{abstract}


\IEEEpeerreviewmaketitle


\printacronyms 

\section{Introduction}
\label{introduction}
Modern \ac{cps} design is a complex process requiring multi-disciplinary expertise. Model and component based engineering techniques are proven approaches based on the traditional idea of "separation of concerns". Each engineering domain relies on their own particular modeling languages, often supported by multiple software tools. These tools provide automation for a wide range of common development tasks, and some existing toolchains provide integrated platforms with many such tools. These platforms provide numerous benefits including version control of system models, automatic tracking and maintenance of generated artifacts, and better cross-domain interaction among others. The \ac{alc} Toolchain \cite{hartsell2019model} is one such platform which assists in the development of \acp{cps} using data-driven development techniques such as machine learning. Any component which uses these techniques is known as a \ac{lec}.

Increasing system complexity and performance demands require close coordination between disciplines as well as more interaction between the supporting software tools. However, most existing platforms only provide automation for particular processes within their respective domains. Outside of these selected processes, the system developer is left with the burden of performing any required interfacing between tools such as data and model transformations, passing of artifacts from one tool to another, and configuration of tools before execution. To address this problem, we introduce a \ac{dsml} for the purpose of modeling \ac{cps} development workflows. This language allows for specification of generic activity graphs where each activity may involve the use of one or more domain-specific tools. The workflow language has executable semantics, hence we implemented both the language and an appropriate execution engine within the \ac{alc} Toolchain. The engine, known as the workflow executor, can automate and monitor the execution of these processes, significantly reducing the burden placed on the developer. 

The remainder of this paper is organized as follows. Section \ref{related_research} presents related research about automation of individual tasks as well as modeling and execution of complete workflows. Next, Section \ref{alc_toolchain} provides a brief introduction to the \ac{alc} Toolchain, the implementation platform for our workflow modeling language. Section \ref{workflow_automation} provides a description of general workflow automation, explains our workflow modeling language, and details how this language was made executable in the \ac{alc} Toolchain. This is followed by a set of illustrative examples in Section \ref{examples}. Finally, we identify areas for future work and give concluding remarks in Sections \ref{future_work} and \ref{conclusion} respectively.

\section{Related Research}
\label{related_research}
Within each engineering discipline, common domain-specific tasks are often automated with appropriate tool support. For example, TensorFlow-eXtended (TFX) \cite{baylor2017tfx} provides an extensible platform for generic \ac{ml} applications. The usual \ac{ml} training process is divided into a 9-step pipeline consisting of smaller individual actions including training data transformation, model learning, and deployment of trained models to production environments. Each step of this pipeline can be customized to fit specific needs before deployment and execution. TFX also automates some common \ac{ml} processes such as hyper-parameter optimization during model training. 

Some platforms take this a step further by automating tasks which have traditionally required multi-disciplinary expertise. The Functional Modeling Compiler (FMC) \cite{canedo2014architectural} is one such approach which uses a high-level model specifying the intended function of a \ac{cps} to generate multiple potential architectures which implement that functionality. These architecture models are composed of component models from multiple engineering domains (eg. electrical, mechanical, chemical) and can be executed in an appropriate simulator. This allows developers to quickly compare alternative system architectures and analyze how design decisions in one engineering domain will affect system-level performance. 

\ac{cps} development platforms facilitate multi-disciplinary development and often integrate multiple tools with varying levels of automation. INTO-CPS \cite{integrated2016larsen} is one such platform built on top of five particular development tools for system modeling, discrete and continuous model analysis, simulation, and test automation. INTO-CPS is intended for use across many \ac{cps} domains and provides examples in several, including automotive, railways, agriculture and building automation. Additionally, INTO-CPS includes recommended development workflows and automates tracking of model provenance and requirements traceability. However, configuration and execution of each operation in the workflow is a manual process. Kulkarni et al. present another framework tailored for smart manufacturing applications \cite{kulkarni2018analytical} which provides tools for the development and refinement of manufacturing models. The authors combine both analytical and data-driven models to facilitate process analysis by domain experts. The framework also integrates and automates execution of several additional libraries including the OpenMDAO Python optimization library \cite{openmdao2019gray}. 

Business Process Model and Notation (BPMN) \cite{omg_bpmn} is an existing modeling standard for representing generic business processes and serves as a general workflow description language. BPMN models typically contain natural language descriptions of each operation in a particular process. Execution engines exist for automating the operations in a BPMN model, but the generic nature of BPMN means very little info required for task execution is contained within the process model. Typically, the individual tools and operations required to complete a process model must be automated independently, then provide a suitable interface to the BPMN execution engine. Similarly, Maier et al. propose a workflow automation system specifically for \acp{cps} based on a new \textit{Project Worker} approach \cite{maier2018efficient}. This approach allows developers to package common, repeatable operations into snippets known as \textit{Engineering Automation Objects}. Additionally, data flowing between workflow operations is identified with \textit{Virtual Addresses} instead of physical addresses which allows data references to be independent of the underlying storage mediums and tool interfaces. The authors provide conceptual examples and identify potential benefits of this approach, but do not implement the proposed workflow automation system in order to validate these benefits. In \cite{mustafiz2012ftg}, Mustafiz et al. introduce the Formalism Transformation Graph (FTG) framework for multi-paradigm modeling. The FTG describes the various \acp{dsml} available and how they are related to one another through model transformations which may be automated or manual. This approach also includes a Process Model (PM) which describes development processes as graph of interconnected model artifacts and activities. Artifacts are used as input to or output from activities, and activities may be one of the model transformations described in the FTG or other external operations.






\section{ALC Toolchain}
\label{alc_toolchain}
\begin{figure*}
    \centering
    \includegraphics[width=1.0\textwidth]{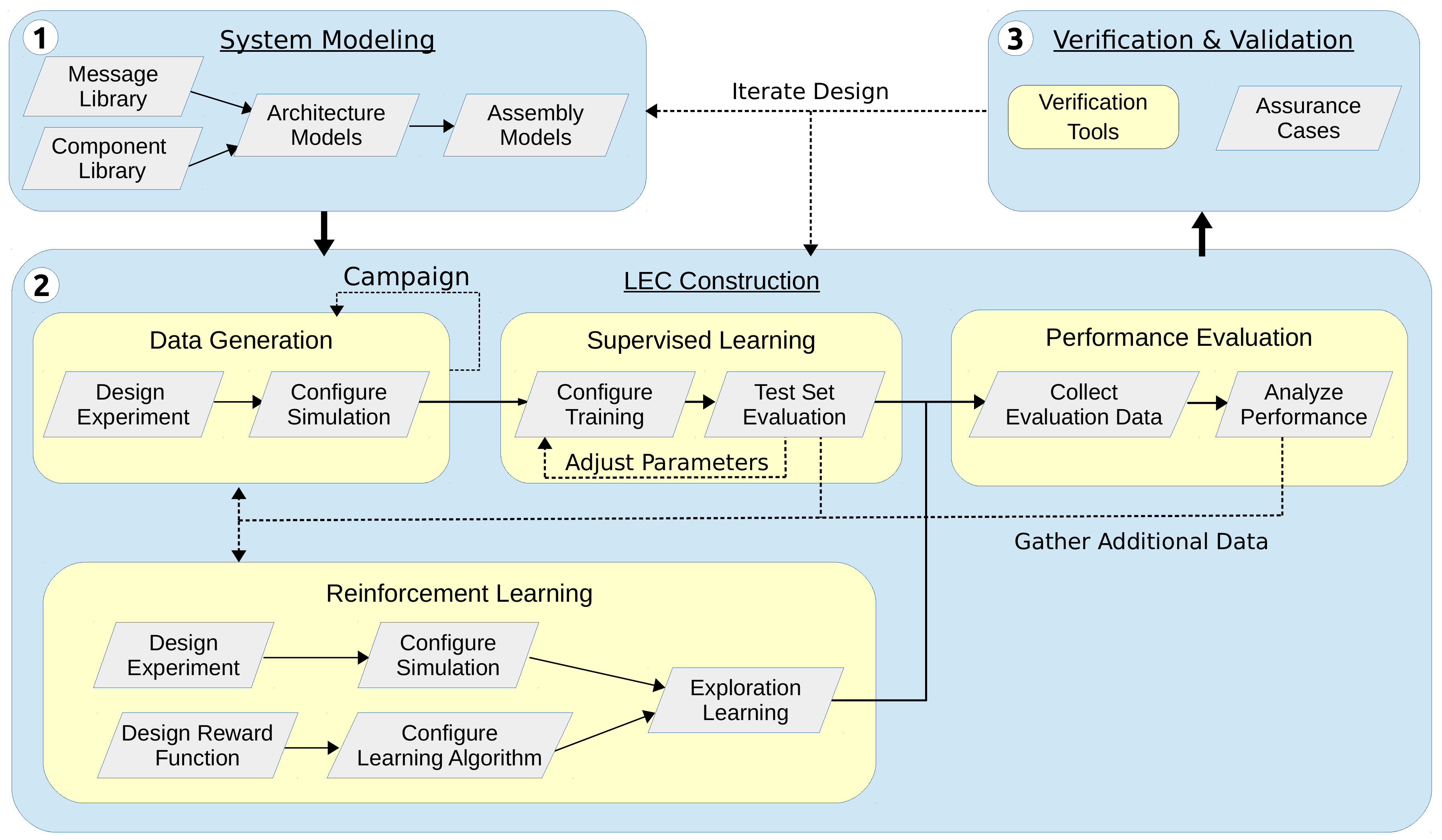}
    \caption{ALC Toolchain development workflow.}
    \label{fig:toolchain_workflow}
\end{figure*}

The \ac{alc} Toolchain supports multi-disciplinary \ac{cps} development by providing an environment with several domain-specific modeling languages and an integrated set of tools. The toolchain is specifically tailored for systems that utilize \acp{lec} and supports development workflows like the one shown in Figure \ref{fig:toolchain_workflow}. In order to promote reproducibility and maintain data provenance, all system models are stored in a version-controlled database and management of all relevant data and generated artifacts is automated. The toolchain is built on the WebGME infrastructure \cite{maroti2014next} which provides a web-based, collaborative modeling environment where changes are automatically and immediately propagated to all active users, not unlike in Google Docs \footnote{\url{https://docs.google.com}}.

The typical development process in the \ac{alc} Toolchain, shown in Figure \ref{fig:toolchain_workflow}, consists of operations divided into two classes: \textit{tasks} and \textit{activities}. \textit{Tasks} are individual actions that require domain-specific knowledge and typically involve the configuration of models or analysis of results. Tasks are shown as grey parallelograms in Figure \ref{fig:toolchain_workflow} with examples including creation of a component library, design of an experiment, or evaluation of \ac{lec} performance. These actions may be performed manually by a system developer or automated with scripts where possible. Once a model has been configured, it can be deployed as an \textit{activity} and executed using domain-specific tools. For example, an \ac{lec} Training model contains all necessary information to train an \acl{ann} using a machine learning library such as TensorFlow \cite{tensorflow2015whitepaper} or PyTorch \cite{pytorch2019paszke}. Activities are shown as yellow, rounded rectangles in Figure \ref{fig:toolchain_workflow} and include common operations such as data generation, \ac{lec} training, and system verification. Additionally, \textit{categories} are used to group related tasks and activities together. System modeling, \ac{lec} construction, and verification \& assurance are the three primary categories in the \ac{alc} Toolchain, shown as blue blocks in Figure \ref{fig:toolchain_workflow}.

\ac{alc} provides automation for execution of individual activities through models which describe the desired operations and contain all the information necessary to initialize and deploy the activity. By default, support is included for several common CPS development tasks, particularly those related to \ac{ml} operations. This automation is designed to be extensible so that new classes of models with different execution semantics can be added in a straightforward manner. Previously, all operations required between execution of activities (eg. passing generated artifacts between models, changing configuration parameters, and starting model execution) were performed manually. These tasks are time consuming to perform and require the user to regularly check if one model execution has completed such that the next execution can be started. The introduction of workflow models into the \ac{alc} Toolchain allows developers to automate these tasks and significantly reduce their workload, as discussed in the following sections.

In the remaining sections of this paper, certain modeling and documentation details have been omitted due to space constraints including the \ac{alc} and workflow \ac{dsml} meta-models, certain details from the provided workflow examples, and the workflow API documentation. However, all of these resources are publicly available on the \ac{alc} Toolchain website\footnote{\url{https://cps-vo.org/group/ALC}}.

    
\section{Workflow Automation}
\label{workflow_automation}
Merriam-Webster defines a workflow as "the sequence of steps involved in moving from the beginning to the end of a working process" \cite{workflow_merriamwebster}. In the context of \ac{cps} development, workflow steps often include complex tasks such as data generation and analysis, performance evaluation, and system-level verification among many others. Typically, operations in a workflow are dependent on results from one or more of the preceding steps. Operations which share the same set of dependencies may be performed in parallel. Additionally, many workflows are iterative processes which are repeated until a predefined threshold is met. To capture these dependencies between operations, workflows are often represented as directed graphs.

Existing CPS development environments usually provide some level of automation for the deployment and execution of individual domain-specific tools, and may enforce certain tool input and output rules to improve the process of interfacing between tools. However, \ac{cps} development workflows usually rely on several interconnected tools with various data transformations between tools. This required interfacing between tools is often automated for a few selected workflows with an appropriate toolchain, or automated on a case-by-case basis in an ad-hoc manner with a general purpose scripting language. While these scripting languages are powerful and generic, creating process automation scripts with them is a time-consuming process which often results in scripts that are fragile in use and not easy to generalize. Additionally, objects and activities within these scripts are not linked to system models or artifacts, and the process of manually synchronizing references as development progresses quickly becomes a significant burden.

To address this problem, we introduce a new \ac{dsml} tailored for the creation and execution of \ac{cps} development workflows and provide one implementation of this language in the \ac{alc} Toolchain. Within \ac{alc}, individual tools are integrated into the toolchain as \textit{activities}. An integrated activity must be able to interpret artifacts used as inputs, often results produced from other activities, and must produce output artifacts which conform to a standard set of rules. The workflow \ac{dsml} allows developers to quickly chain multiple activities together to model their typical development processes in a structured manner and deploy these processes for execution. This approach is intended to bring the benefits of model-based engineering to CPS workflow automation and is explained in more detail in the following sections.

\subsection{Modeling Language}
\begin{figure}
    \centering
    \includegraphics[width=1.0\columnwidth]{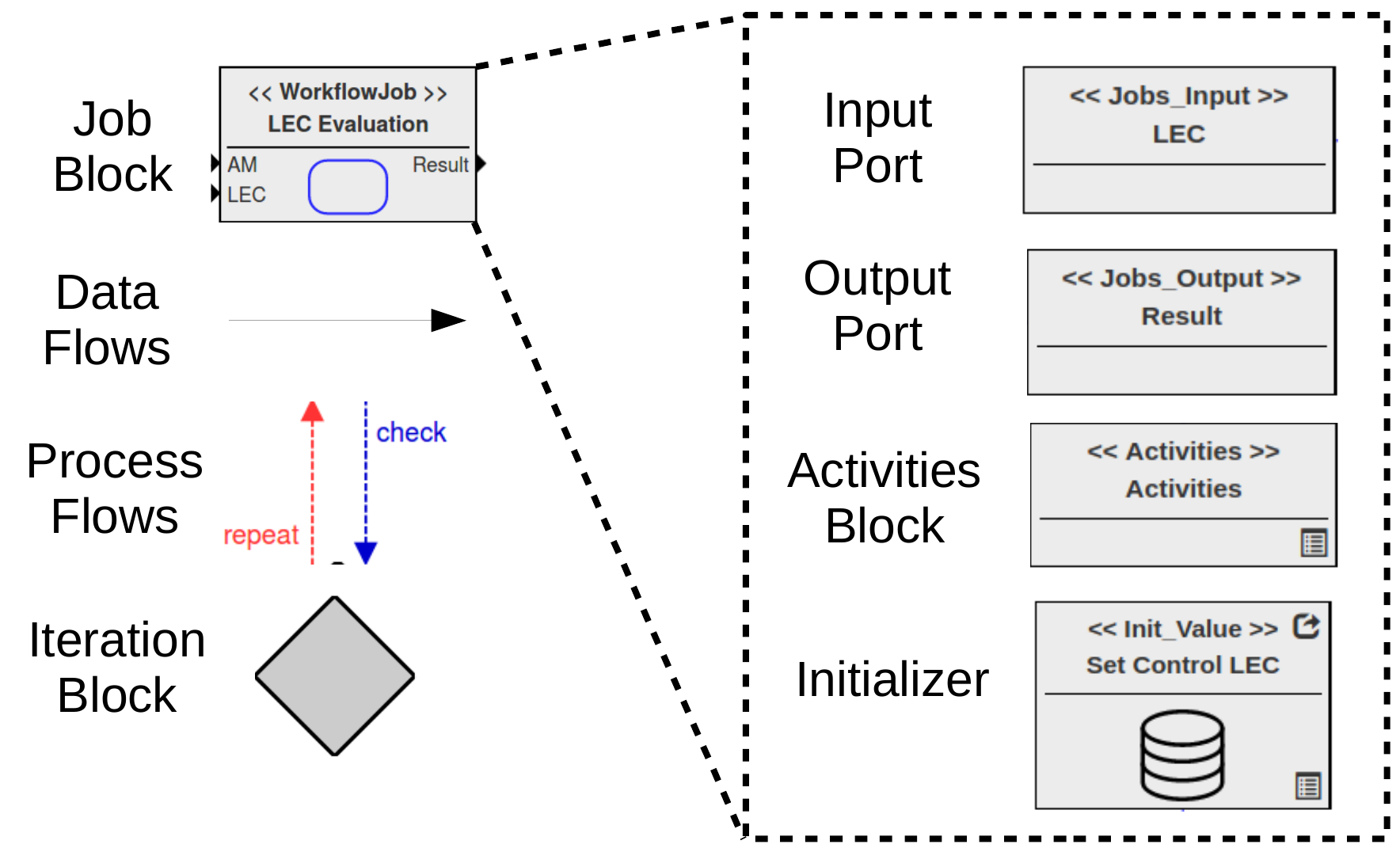}
    \caption{Components of the Workflow modeling language.}
    \label{fig:model_legend}
\end{figure}

The Workflow \ac{dsml} within \ac{alc} consists of the eight primary modeling objects shown in Figure \ref{fig:model_legend}. Construction of a workflow model starts with the creation of a \textit{job} block, which describes one step in the workflow process and contains exactly one \textit{activities} block. Within \ac{alc}, any executable model can be considered an activity, and execution typically involves the use of domain-specific tools which operate on the input model and produce a result. The activities block within a workflow job specifies which models should be executed during this job, and contains links to one or more existing activity models. Each individual activity model usually needs to be initialized before execution which can be done with an appropriately configured \textit{initializer} block. Activity initialization is often dependent on the results of previous jobs in the workflow, which can be accessed through one or more \textit{input ports}. The results of each activity contained within a job can be made available to later jobs through \textit{output ports}. Finally, \textit{iteration} blocks can be used to guide the execution flow of the model by specifying where and how the flow should be interrupted to perform additional iterations. These blocks contain a script which examines any of the outputs from previous jobs, determines if additional iterations of the workflow should be performed, and updates the activity models as needed for the next iteration.

\begin{figure*}
    \centering
    \includegraphics[width=1.0\textwidth]{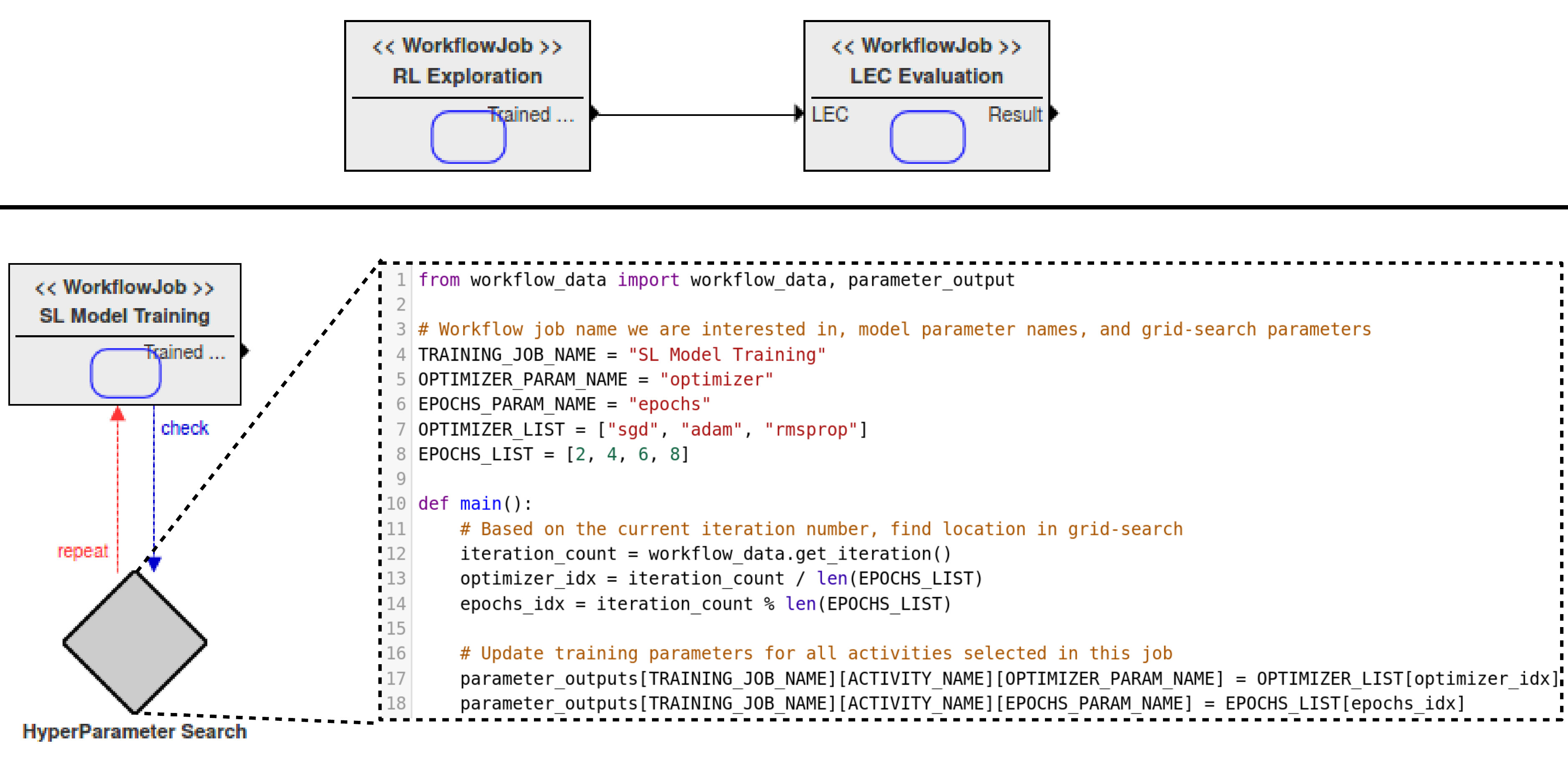}
    \caption{Workflow models for example 1 (top) and 2 (bottom). The "HyperParameter Search" block in example 2 contains a script, shown with a dotted outline in the figure.}
    \label{fig:workflow_examples_1_and_2}
\end{figure*}

A complete workflow model, like the examples shown in Figures \ref{fig:workflow_examples_1_and_2} and \ref{fig:workflow_example_3}, consists of one or more \textit{jobs} connected in a directed graph where each edge is one of two types. \textit{Data flow} edges are depicted as black lines and represent the flow of generated artifacts from one workflow job into another job. These edges are always directed from an output port to an input port. \textit{Process flow} edges direct the execution flow of the workflow and may be one of two types. \textit{Check} edges, shown as blue lines, indicate where the normal flow of the workflow should be interrupted to check if any iteration is needed. If the script contained within the \textit{iteration} block, shown as a grey diamond, indicates that iteration is desired then \textit{repeat} edges, shown as red lines, indicate where the next iteration should begin. This structured, model-based approach to workflow creation allows developers to quickly automate their usual development processes in a generic, reusable, and extensible manner. The example model in Figure \ref{fig:workflow_example_3} corresponds to one specific instance of the general \textit{LEC Construction} phase of the \ac{alc} development workflow shown in Figure \ref{fig:toolchain_workflow} and is discussed in more detail in Section \ref{example3}. A meta-model for the workflow \ac{dsml} described above was created in the \ac{alc} Toolchain using the tools provided by the WebGME environment. Due to space constraints, the meta-model has been omitted here, but is available to explore on the \ac{alc} Toolchain website. Accessing this model requires registering a user account, clicking the large "LAUNCH" button, then selecting the "ALC\_Meta" project.

\subsection{Implementation}
We have created a \textit{Workflow Executor} within the \ac{alc} Toolchain which handles deployment, execution, and data management for workflow models. Scripts contained in \textit{iteration} blocks which guide looping within the workflow are specified in the Python\footnote{\url{https://www.python.org/}} programming language and are invoked on each iteration of their corresponding loop. The user-specified behavioral code within these scripts can be arbitrarily complex. An API is provided for interacting with models in the \ac{alc} Toolchain including ability to: 

\begin{itemize}
    \item Access all data produced by previously executed jobs in the workflow.
    \item Persist selected data across loop iterations.
    \item Update any model parameters for future activities in the workflow.
\end{itemize}

The workflow executor is implemented on top of the \textit{Gradle}\footnote{\url{https://gradle.org/}} build automation system. Gradle is intended for software build tasks such as compilation, packaging, and deployment. However, it provides a Kotlin\footnote{\url{https://kotlinlang.org/}} based build script language and is easily extensible for other more specific purposes. In particular, Gradle provides well-defined, easy to use interfaces which are can be integrated with other platforms including the WebGME environment that the \ac{alc} Toolchain is built upon. When a workflow model is executed, the executor first extracts all relevant execution information from the model and stores this information in a JavaScript Object Notation (JSON) file. This JSON file is then interpreted by the Gradle build script, which generates the necessary task graph to carry out the workflow dynamically based on the constraints and dependencies specified JSON. The Gradle engine is well-suited to executing the workflow as it provides built-in functionality for:

\begin{itemize}
    \item dependencies to be expressed between the tasks, ensuring that they are executed in the proper order
    \item detection, reporting of, and termination of the workflow due to any errors during execution
    \item determining if a task should be skipped, e.g. in the case that a previous execution of the workflow already executed the task, and its output need not be regenerated.
\end{itemize}

Gradle is only capable of generating a \ac{dag} of tasks, i.e. no loops can be in the task dependency graph. Therefore, to implement looping in the Workflow \ac{dsml}, a separate process is executed that contains a simple loop. Each iteration of the loop executes a Gradle process to build and execute a workflow task \ac{dag} for that specific iteration. The loop carries information forward from the previous iteration to the next so the appropriate \ac{dag} for the workflow is executed on each iteration. 

\begin{figure}
    \centering
    \includegraphics[width=1.0\columnwidth]{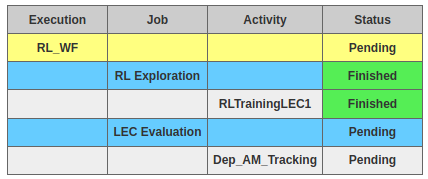}
    \caption{Workflow status visualization.}
    \label{fig:status_viz}
\end{figure}

Once a workflow execution begins, the executor tracks the progress of each job in the workflow and provides this information back to the user with a status table like the one shown in Figure \ref{fig:status_viz}. If a workflow job encounters an error during execution, the status table will be updated to reflect this. Once an error has been identified and fixed, then the workflow can be resumed from the point where the failure occurred. This way, none of the preceding jobs which finished successfully before the error will be re-executed.

\section{Examples}
\label{examples}

The following sections present several examples showing how common workflows used in \ac{alc} can be automated with appropriate workflow models. All examples are in context of an \ac{uuv} tasked with following a pipeline on the seafloor using two \acp{lec} along with several non-LEC components. Both \acp{lec} are neural network based, but use different training algorithms provided by different machine learning libraries. The Perception \ac{lec} performs image segmentation on input sonar images and produces a segmented image where every pixel in the original image has been assigned to an object class (eg. pipeline, seafloor, obstacle). This is accomplished with a \ac{cnn} based on the SegNet \cite{segnet2015badrinarayana} architecture and trained with a supervised learning approach using the PyTorch \cite{pytorch2019paszke} library. The Control \ac{lec} must then use this information, as well as additional vehicle state information, to output appropriate control commands for the \ac{uuv}. The control \ac{lec} is a two-layer, feed-forward network trained using reinforcement learning techniques provided by the Tensorforce \cite{tensorforce} library. For a detailed introduction to supervised and reinforcement learning techniques see \cite{Goodfellow-et-al-2016} and \cite{sutton2018reinforcement} respectively. Additionally, each of the example models shown in the following sections can be explored in full detail on the \ac{alc} Toolchain website.

\subsection{Reinforcement Learning \& Evaluation}
\label{example1}

The first example workflow involves training the control \ac{lec} over many simulated, randomly-generated scenarios known as \textit{episodes}. During each episode, the \ac{lec} explores the possible action space while using the provided reward function to observe which control actions produce desirable results for the \ac{uuv} system. Network weights are updated accordingly, and this process continues for a set number of episodes. Once the training process is completed, the \ac{lec} can be deployed back to the simulator in a non-learning mode (ie. network weights are no longer updated) for system-level evaluation. During this job, the control \ac{lec} is evaluated based on system-level performance metrics of the \ac{uuv} instead of the component-level metric provided by the reward function. Evaluation jobs are often configured as a \textit{campaign} where the same evaluation procedure is repeated in a variety of scenarios which can be provided as a specific set or generated with an appropriate scenario description language such as Scenic \cite{scenic2018fremont}.

This training and evaluation process can easily be modeled with our workflow modeling language as shown in the top of Figure \ref{fig:workflow_examples_1_and_2}. The training process is captured in the "RL Exploration" block which provides a trained \ac{lec} output. This output is fed into the "LEC Evaluation" block and used to initialize the evaluation models. The result of each evaluation activity is also made available, but is not used in this workflow. While the \ac{alc} Toolchain provides automation of each individual activity, previously no automation between activities was available and all data passing and configuration between models required manual configuration. This particular workflow model was created by an experienced \ac{alc} user in about 15 minutes, but may take hours or days to execute depending on the number of learning episodes, evaluation scenarios, and available computational resources. Additionally, the status visualizer for this example is shown in Figure \ref{fig:status_viz}. In this case, the "RL Exploration" job had completed successfully, indicated by the green "Finished" status, and the "LEC Evaluation" job was pending deployment to an execution server.

\subsection{Hyperparameter Optimization}
\label{example2}

The next example involves automating hyperparameter optimization for the supervised training of the perception \ac{lec}, a common process in the machine learning field. The goal of this process is to find a set of training configuration parameters, known as \textit{hyperparameters}, which results in an \ac{lec} with the minimal total loss when evaluated against a particular set of test data. While many common loss functions are available, we use a standard mean-squared-error loss function in this example. Common optimization search algorithms include grid search, random search \cite{random2012bergstra}, and Bayesian optimization \cite{practical2012snoek}. For the purpose of this example we automate a simple grid search over two particular hyperparameters, gradient-descent optimizer and number of epochs, but more advanced search algorithms can be automated in a similar fashion.

\begin{figure}
    \centering
    \includegraphics[width=1.0\columnwidth]{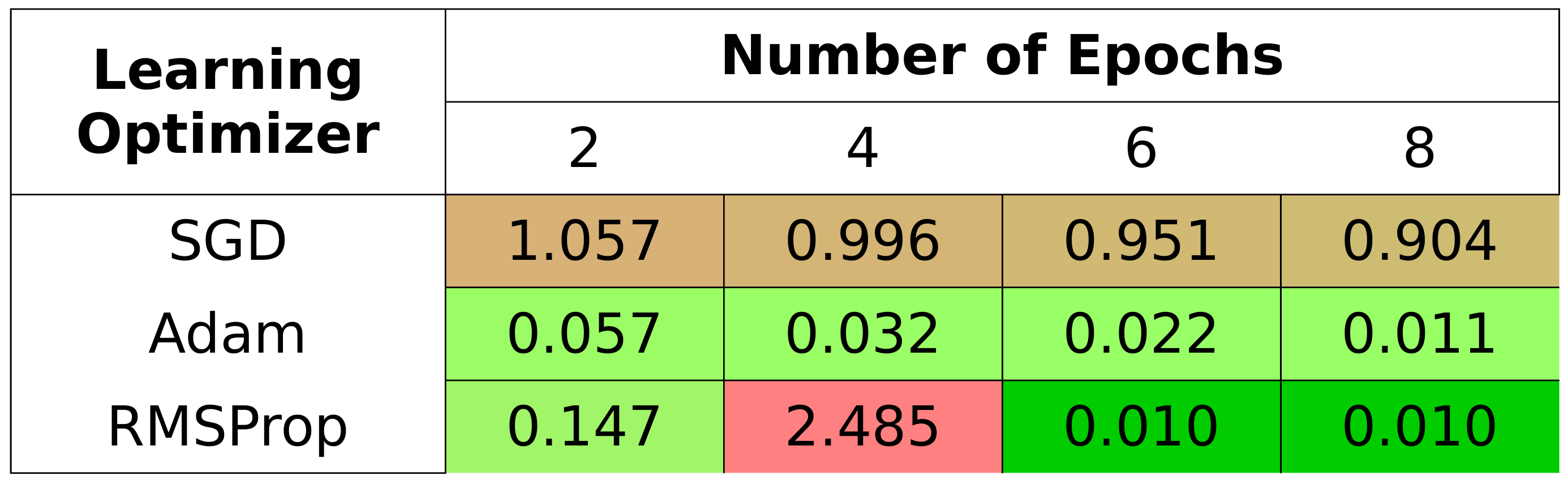}
    \caption{Perception LEC loss after training with various hyperparameters. Large loss values are colored red while small loss values are colored green.}
    \label{fig:hyperparam_results}
\end{figure}

Modeling the hyperparameter optimization process requires introducing the \textit{iteration} concept as shown in the bottom left of Figure \ref{fig:workflow_examples_1_and_2}. In this workflow, there is only one job named "SL Model Training" which initializes the training activity model with labeled data sets before executing the training process for the perception \ac{lec}. The trained LEC is produced as an output, but is unused in this workflow. This job is then connected to a "Hyperparameter Search" iteration block, shown in an exploded view, containing the script which implements the iteration logic. In this case, each iteration will repeat the training process with one of three learning optimizers: Stochastic Gradient Descent (SGD) \cite{robbins1951stochastic}, Adam \cite{kingma2014adam}, or RMSprop \cite{hintonRMSProp}. Additionally, each optimizer will be run for 2, 4, 6, and 8 training epochs to evaluate how quickly each optimizer converges to the minimal loss. This results in a 3 by 4 search grid for a total of 12 training iterations, and the final loss value of each combination of hyperparameters is shown in Figure \ref{fig:hyperparam_results}. In this case, both the RMSprop and Adam optimizers converge much faster than the standard SGD technique, but the RMSprop optimizer has a large jump in error rate when trained for 4 epochs. This could be due to other training parameters which were not considered in this test, such as a learning rate that is too large. Increasing the number of learning parameters used in our workflow grid search could help confirm this hypothesis, but was out of scope for this paper. Once all 12 iterations are completed, the workflow will terminate automatically. Similarly to the previous model, this workflow model required less than an hour of developer time to construct, but may require many computational hours to execute all 12 \ac{lec} training cycles. Without the workflow \ac{dsml} and executor, the developer would have to regularly monitor for completion of each training execution, then manually update the training activity model with the next set of network hyperparameters and launch the next iteration of the grid search. Manual monitoring for job completion is tedious and typically results in available computing hardware sitting idle until the developer notices job completion. Additionally, manually updating model parameters and references for this relatively simple example is a quick, straight-forward process, but quickly becomes difficult as the complexity of the workflow increases.

\subsection{ALC Development Workflow}
\label{example3}

\begin{figure*}
    \centering
    \includegraphics[width=1.00\textwidth]{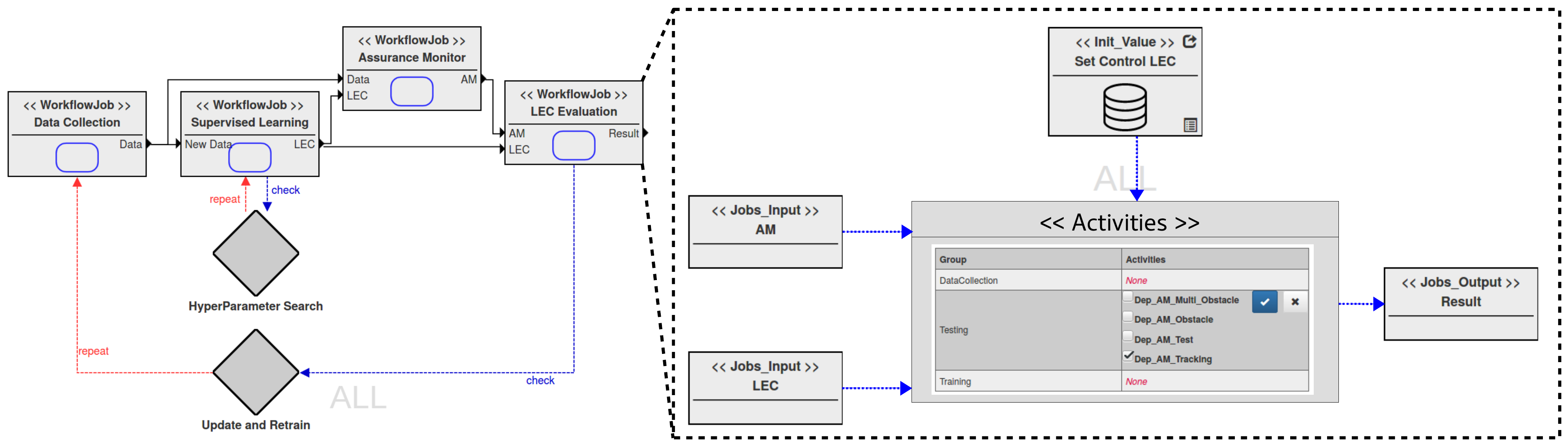}
    \caption{Workflow model for example 3. The "LEC Evaluation" workflow job includes an exploded view of the internal model.}
    \label{fig:workflow_example_3}
\end{figure*}

The last example demonstrates automation for the full "LEC Construction" portion of the \ac{alc} development workflow shown in Figure \ref{fig:toolchain_workflow}. This process consists of generating labeled training data through simulation of the \ac{uuv} system in a variety of scenarios. For this system, variations between scenarios include different configurations of the pipeline on the seafloor, randomized starting location of the \ac{uuv}, and randomly generated obstacles in the path of the vehicle among others. The collected data can then be passed to a supervised learning model in order to train the perception \ac{lec}. At this point, the developer may chose to perform a hyperparameter search to optimize the performance of the \ac{lec} against an evaluation data set. Once an optimal training configuration has been identified, both the training data set and the trained \ac{lec} are sent to an assurance monitor training job. Assurance monitors are software components which act in parallel to an \ac{lec} and provide a confidence metric for each output produced by that \ac{lec}. This metric is based on the similarity between a new, previously unseen input and all inputs contained in the data set used to train the particular \ac{lec}. This technique is known as Inductive Conformal Prediction \cite{papadopoulos2008inductive}, and there are a number of algorithms for efficiently computing confidence metrics. In this example, we use one particular algorithm \cite{realtime2020cai} based on Support Vector Data Description \cite{tax2004support}. Finally, an evaluation model can be configured to use the trained perception \ac{lec} and corresponding assurance monitor as part of the complete \ac{uuv} system. If the \ac{lec} shows unsatisfactory performance during a particular evaluation scenario, then the developer may chose to add this scenario to the original data generation step and repeat the process.

Figure \ref{fig:workflow_example_3} shows how this full LEC construction workflow can be modeled. As with the previous two examples, each type of activity is contained in a workflow job block with appropriate input and output ports to control the flow of generated artifacts. The last job in the workflow, LEC Evaluation, is depicted in an exploded view showing how individual activity models can be selected for this job. In this case, only a single evaluation model named "Dep\_AM\_Tracking" has been selected to run. However, this model is configured as a campaign so that several simulated scenarios will be run to evaluate the trained \ac{lec}. Additionally, this workflow model contains two loops: an inner loop for hyperparameter optimization and an outer loop for iterating the entire process with additional training scenarios. The inner loop follows the same procedure outlined in Section \ref{example2} to optimize the \ac{lec} training procedure. The outer loop contains a script which identifies any evaluation scenarios which resulted in unsatisfactory performance and updates the first workflow job, "Data Collection", to include these specific scenarios in the training data set. Also note that the assurance monitor training job is dependent upon the trained \ac{lec} produced by the hyperparameter search loop. Because of this dependency, the assurance monitor job will not run until after the entire hyperparameter search has completed. Since the hyperparameter search is repeated in full on every iteration of the outer loop, this workflow typically executes many computationally intensive \ac{lec} training activities. As with the previous examples, this workflow model was constructed and deployed in less than an hour of developer time, but required several computational hours to complete.

\section{Future Work}
\label{future_work}
We have identified several opportunities for future improvements in both the workflow language itself as well as the implementation of the language within the \ac{alc} Toolchain. First, the workflow language does not allow general branching operations, which limits their expressiveness. While looping is supported, this can be viewed as a special case of more generic branching. Additionally, the language does not allow for hierarchical construction of workflows. That is, one workflow model cannot contain another workflow model. Hierarchical models can be useful in many scenarios and can promote model reuse. For instance, the hyperparameter optimization loop in the final example presented in Section \ref{example3} could have been replaced by the workflow model shown in the bottom of Figure \ref{fig:workflow_examples_1_and_2}. Due to this current limitation, the optimization loop model had to be recreated for both examples. We intend to add both general branching and hierarchical construction features to the workflow \ac{dsml}.

Within the \ac{alc} Toolchain, some support is provided for the debugging of activities including the ability to quickly view all execution logs and perform interactive executions with Jupyter notebooks\footnote{\url{https://jupyter.org/}}. However, debugging capabilities for the workflow executor are limited and require understanding of the executor internals. In particular, there is no straightforward way of debugging the user-provided \textit{iteration} scripts which guide the execution of the workflow. We plan to add additional debugging capabilities to allow typical developers to troubleshoot models without detailed understanding of the executor internals. Next, our current implementation of the Workflow Executor does not support any parallel execution of job iterations within loops. For example, each training session of the hyperparameter optimization described in Section \ref{example2} could be executed in parallel since the execution of one iteration is not dependent on the results from the previous iteration. However, this type of parallelization would not apply to loops with dependencies between each iteration such as the outer loop of the development cycle described in Section \ref{example3}. Additionally, a primary goal of the \ac{alc} Toolchain is to assist developers in providing system-level assurance for \acp{cps}. With this in mind, we intend to integrate our workflow models more tightly with existing tools for construction and evaluation of assurance cases. We expect this would reduce both the burden of manually maintaining these assurance cases and the time required to identify system-level properties after design iterations. Finally, the current workflow executor is able to catch error conditions and report them back to the user. However, these errors must be corrected manually as the executor does not have any fault mitigation or recovery capabilities. It is often the case that a developer will not be able to address such issues until hours or days after the error occurred, resulting in wasted computer time. We plan to explore this functionality in future work to alleviate this problem.

\section{Conclusion}
\label{conclusion}
Modern \ac{cps} design requires cross-discipline expertise and involves many heterogeneous system models. These models often include domain-specific tooling for various tasks such as performance analysis, code generation, and system deployment among others. Automation of these tools has typically been limited to execution of individual tasks or to a limited set of common discipline processes. To address this problem, we have introduced a \ac{dsml} for describing \ac{cps} workflows and provided an executive engine for automated execution of these workflows. This, along with the existing automation capabilities provided by the \ac{alc} toolchain, helps to bring the full benefits of model-based engineering to workflow automation.

\balance

\section*{Acknowledgement}
This work was supported by DARPA and the Air Force Research Laboratory. Any opinions, findings, and conclusions or recommendations expressed in this material are those of the author(s) and do not necessarily reflect the views of DARPA or AFRL.

\bibliographystyle{IEEEtran}
\bibliography{references}
\end{document}